\documentclass[draftcls,onecolumn,12pt]{IEEEtran}

\usepackage{cite}
\usepackage{amsmath,amssymb,amsfonts,amsthm}
\usepackage{algorithmic}
\usepackage{graphicx}
\usepackage{textcomp}                                          
\usepackage{multicol}         
\usepackage{caption}
\usepackage{amssymb,amsmath,dsfont}
\usepackage{inputenc}
\usepackage{bbm}
\usepackage{comment}
\usepackage{cite}
\usepackage{color}
\usepackage{multirow}
\usepackage{subcaption}
\usepackage{array}
\usepackage{algorithm}
\usepackage{color,soul}
\usepackage{blindtext}

\makeatletter
\newcommand{\mathleft}{\@fleqntrue\@mathmargin0pt}
\newcommand{\mathcenter}{\@fleqnfalse}
\makeatother

\begin{document}
\title{SWIPT-based Real-Time Mobile Computing Systems: A Stochastic Geometry Perspective }

\author{Ayse~Ipek~Akin,  \IEEEmembership{Member,~IEEE}
	Nafiseh Janatian, \IEEEmembership{Member,~IEEE}
        Ivan~Stupia, \IEEEmembership{Member,~IEEE}
        and~Luc~Vandendorpe,~\IEEEmembership{Fellow,~IEEE}
\thanks{A. I. Akin, N. Janatian, I. Stupia and L. Vandendorpe are with the Institute of Information and Communication Technologies, Electronics and Applied Mathematics (ICTEAM), Universit\'e catholique de Louvain, Louvain la Neuve, Belgium.}}

\maketitle

\begin{abstract}
	
Driven by the Internet of Things vision, recent years have seen the rise of new horizons for the wireless ecosystem in which a very large number of mobile low power devices interact to run sophisticated applications. The main hindrance to the massive deployment of low power nodes is most probably the prohibitive maintenance cost of battery replacement and the ecotoxicity of the battery production/end-of-life. An emerging research direction to avoid battery replacement is the combination of radio frequency energy harvesting and mobile computing (MC). In this paper, we propose the use of simultaneous information and power transfer (SWIPT) to control the distributed computation process while delivering power to perform the computation tasks requested. A real-time MC system is considered, meaning that the trade-off between the information rate and the energy harvested must be carefully chosen to guarantee that the CPU may perform tasks of given complexity before receiving a new control signal. In order to provide a system-level perspective on the performance of SWIPT-MC networks, we propose a mathematical framework based on stochastic geometry to characterise the rate-energy trade-off of the system. The resulting achievable performance region is then put in relation with the CPU energy consumption to investigate the operating conditions of real-time computing systems. Finally, numerical results illustrate the joint effect of the network densification and the propagation environment on the optimisation of the CPU usage.

\end{abstract}


\IEEEpeerreviewmaketitle

\section{Introduction}
In the recent years, the saturation of the smartphone market penetration fostered the development of new applications to connect human beings with smart objects. The collection of sensors, actuators, algorithms and connectivity enabling complex machine-to-machine and human-to-machine interactions is mostly referred to as \emph{the internet of things} (IoT). The success of this technological paradigm among both academic and industrial players led to new challenges about the possible limitations in energy storage and computing capacity for the \emph{mobile low power devices} (LPDs) that will populate the future wireless ecosystem. Currently, researchers and engineers are striving to lay the foundations of a technological breakthrough towards a new generation of LPDs with enhanced real-time processing capability that would support the IoT vision. In this sense, the appealing solutions proposed so far delineate a new framework for the design and the optimisation of wireless networks in which computation and traditional communication aspects are merged into a unified perspective \cite{barbarossa2014}. The consequence is a new network formalism in which the final objective of the design process is not the maximisation of the transferable information \emph{per se}, but rather the maximisation of the information instrumental in guaranteeing an optimal use of the limited computation resources available at the LPDs \cite{li2018fundamental}.

Following this line of thought, it appears that the implementation of such mobile computing (MC) systems is confronted with a double bottleneck: \emph{i)} the restrictions on the amount of energy available to perform sophisticated computation tasks, and \emph{ii)} the limited capacity of the radio links. A potential solution to overcome this double obstacle consists in a joint exploitation of the communication and computation facilities to decompose demanding tasks and execute them across the whole network,\cite{barbarossa2014,li2018fundamental,mao2017}. Within this context, an enticing emerging research direction is the integration of wireless power transfer (WPT) and MC \cite{you2016energy,mao2016dynamic}. In WPT-MC systems, the computation power can be obtained through radio frequency energy harvesting (RFEH) to prolong the battery life of the LPDs \cite{bi2017computation,wang2017joint}. For example, in \cite{bi2017computation}, the authors analysed a multi-user wireless powered MC system with computation offloading capability and proposed a mode selection scheme to maximise the sum of the computation rates. Concurrently, the authors of \cite{wang2017joint} studied an optimal resource allocation scheme to minimise the total energy consumption under computation latency constraints. Lately, in \cite{janatian2018optimal}, simultaneous wireless information and power transfer (SWIPT) was proposed to jointly provide power and information to LPDs in an MC system.

\subsection{Motivations}
A crucial aspect to be considered in designing SWIPT systems is the attenuation of radio waves with distance, which is known as the path-loss. Path-loss is particularly detrimental for energy transfer and may even question its relevance!  As an example, in \cite{lu2015b}, the authors show that an isotropic wireless energy source delivering $4$ Watts in the $900$MHz band would enable a device at $15$m to harvest only $5.5\mu$W in free space. Moreover, it is worth remarking that the public safety norms would inhibit the possibility of an increased received signal strength via a further increase of the radiated power. It follows that the only possible strategy to counteract path-loss is to densely deploy wireless energy sources, thus enabling lower transmit power requirements and increased levels of the energy delivered to the LPDs. Network densification has already been extensively studied as one of the key ingredients for the development of the 5G infrastructure, \cite{thompson2014,hwang2013}. One of the main results of these studies is that traditional network modelling approaches may fail in describing dense networks. In order to deal with this problem, stochastic geometry (SG) has been introduced as a random and spatial model to analyse the performance of dense networks \cite{baccelli2009}. A mathematical framework for stochastic geometry analysis of SWIPT systems was originally proposed in \cite{lu2015stochastic} and subsequently extended in \cite{thanh2016mimo,di2017system,lam2016system} for MIMO SWIPT networks. However, all those studies consider outdoor cellular networks and propagation models based on distance-based path-loss functions only, thus failing in describing the correlated structure of the signal blockage due to buildings and walls. In this respect, it is worth mentioning that the nonlinearity of the harvesting devices does not allow to efficiently harvest energy from weak signals (i.e. less than -30 dBm of received power). This technological constraint, together with the limitations on RF emissions for guaranteeing public safety, suggests that the deployment of SWIPT transmitters must be denser than that of traditional access points in wireless networks, thus making unlikely the deployment of SWIPT systems in outdoor scenarios. For this reason, a generative model relying on stochastic geometry has been developed to analyse the shadowing effect for in-building systems in \cite{zhang2015indoor}.  More recently, a similar model has been adopted in \cite{akin2018} for the analysis of the rate-energy trade-off in dense MIMO SWIPT indoor networks.

 Differently from the conventional communication systems, in SWIPT-MC systems, the spatial distribution of the network nodes would play a major role, not only on the performance of the communication links, but also in determining the computation capacity of the LPD nodes. This is mainly due to the fact that the complexity of the allowed computation tasks is limited by the amount of harvestable power \cite{bi2018computation}. In particular, in SWIPT-MC systems, the ability of an LPD to perform given tasks is reflected on specific rate-energy trade-offs: more energy is furnished to the LPDs for the computation tasks, more likely the multi-user interference will weaken the communication capacity, \cite{akin2018}. To the best of our knowledge, a stochastic geometry analysis that investigates the interplay between link capacity, energy transferred and complexity of the computation tasks at a network level is still missing in the literature. The main objective of this paper is to close this gap.

\subsection{Contributions}

In this paper we develop a methodology to analyse an MC system in which SWIPT is used to simultaneously deliver power and control information to the LPDs. We assume that the MC system supports a given set of services and that each service is associated with a list of computation tasks to be performed by multiple LPDs in a distributed fashion. In order to minimize the global energy consumption of the LPDs, we assume that each low-power node is specifically designed to allow the execution of computation tasks associated with only a subset of services. Each SWIPT transmitter is allocated to only one service during a single time slot and it transmits the control information to the LPDs associated with the same service. The control signal is used to schedule specific tasks at the LPD, out of the set associated with the relevant service. The original contributions of this paper can be summarised as follows:
\begin{itemize}

\item	Capitalizing on SWIPT, we propose an original model in which the same waveform is used to deliver power and control signals. Thanks to this, a computation service can be split into elementary computation tasks to be distributed across disparate LPD nodes. This procedure has the advantage of avoiding the execution of decision-making algorithms at the LPDs that might consume more power than the one obtained through harvesting.

\item	To enable a simplified design of the LPD node and further reduce its energy consumption, we propose a real time mobile computing system, so that costly queue and latency management at the LPDs can be avoided.

\item In order to optimize the energy efficiency of the LPDs, engineers often opt for designs that are specifically tailored to particular applications. Hence, in this work, we consider that each LPD supports only a subset of services. Moreover, we assume that each PH can be used by only one service provider at the time. This translates to a model in which a generic PH and a typical LPD are associated with the same service with a given probability. Hence,  the SG approach originally proposed in \cite{akin2018} has been generalized to enable the analysis of the rate-energy trade-off on a per-service basis.

\item	We provide the basic tools to understand the connection between the rate-energy trade-off and the computation capacity of the LPDs. In order to achieve the computation capacity, it is assumed that the central processing unit (CPU) clock frequencies are optimised to process the local data with the minimum energy consumption.

\item The outage probability for the computation capacity of the SWIPT-MC system in an indoor environment is quantified. It is revealed that the best CPU utilisation is obtained for more complex tasks performed at a lower rate rather than performing less complex tasks at a higher rate.

\item Unlike other currently available analytical frameworks, the impact of network densification on the MC system's performance is revealed, showing that the optimal level of densification actually depends on the targeted task complexity.

\item The role played by the topology of the venue, and in particular the frequency of the blockage objects (e.g. walls), is also contemplated in our framework. This allows a macro-level investigation of the interactions between the SWIPT-based MC system and the venue in which it is deployed. 
\end{itemize}
Finally, it is worth remarking that in a real-time MC system, the computation capability of the LPD must rely on the energy harvested during a single transmission slot. To achieve this, the computation tasks are modelled through the number of logical operations per bit that must be performed and the size of the local data to be processed. To the best of our knowledge this is the first paper that proposes an SG approach for the study of real-time SWIPT-MC systems.

\subsection{Notations} 
$\Gamma(\cdot)$ is the Gamma function, \cite[Eq.8.310.1]{gradshteyn2014table}. $\Gamma(\cdot, \cdot)$ denotes the upper-incomplete Gamma function \cite[Eq.8.350.2]{gradshteyn2014table}. $\operatorname{Im}\{\cdot\}$ is the imaginary part,  $_pF_q(a_1;\dots,a_p;b_1,\dots,b_q;\cdot)$ is the generalized hypergeometric function \cite[Eq.9.14.1]{gradshteyn2014table} and $\mathcal{H}(\cdot)$ denotes the Heaviside function.

\section{System Model}
 
In this section, we model a MIMO SWIPT-MC system operating in an indoor environment. The objective of the MC system is to provide a set of services $\mathcal{S}$, where the attributes of a generic service $s \in \mathcal{S}$ are a collection of $V$ computation tasks $\mathcal{C}_s = \{c_{s,1},c_{s,2}, \cdots, c_{s,V}\}$. Each computation task processes $N$ bits representing contextual information locally stored at the LPD. Examples of contextual information can be sensed data, information on the activity of a human user, mobility patterns, etc. The SWIPT waveform received is used to simultaneously decode the control signal that will select the computation tasks to be executed and extract the power needed to perform those tasks. According to this model, instead of considering the signal propagation as the sole restriction, we recognise that the SWIPT signals are sent with the purpose of enabling the execution of computation tasks. Hence, besides the signal propagation model, we also present a practicable model for the CPU energy consumption at the LPD. 

\subsection{Network model}

The network infrastructure is composed of a set of SWIPT transmitters, also referred to as power heads (PHs), randomly located over a finite region. 
We assume that the PHs have been deployed following a homogeneous Poisson point process (PPP) distribution, $\Psi$, with density $\lambda_{PH}$. At each time slot, a generic PH is associated with the service $s$ with probability $q_s$. We denote with $\mathcal{P}_s$ the subset of PHs allocated to the service $s$. During an initial discovery phase, the LPDs broadcast a packet containing a message indicating their associated services. Then, each LPD establishes a communication link with the PH $p \in \mathcal{P}_s$ ensuring the minimum average signal attenuation, where $s$ is the service associated with that LPD.  We also assume that all the PHs share the same radio resources, so that all the PHs but $p$ are considered as interferers for the information transfer process. The set of interferers for the $l$th LPD is denoted by $\Psi^{(\,\backslash l)}$. An illustration of the SWIPT-MC system is provided in Fig.1.

\begin{figure}[!t]
	\centering
	\includegraphics[width=3.5in]{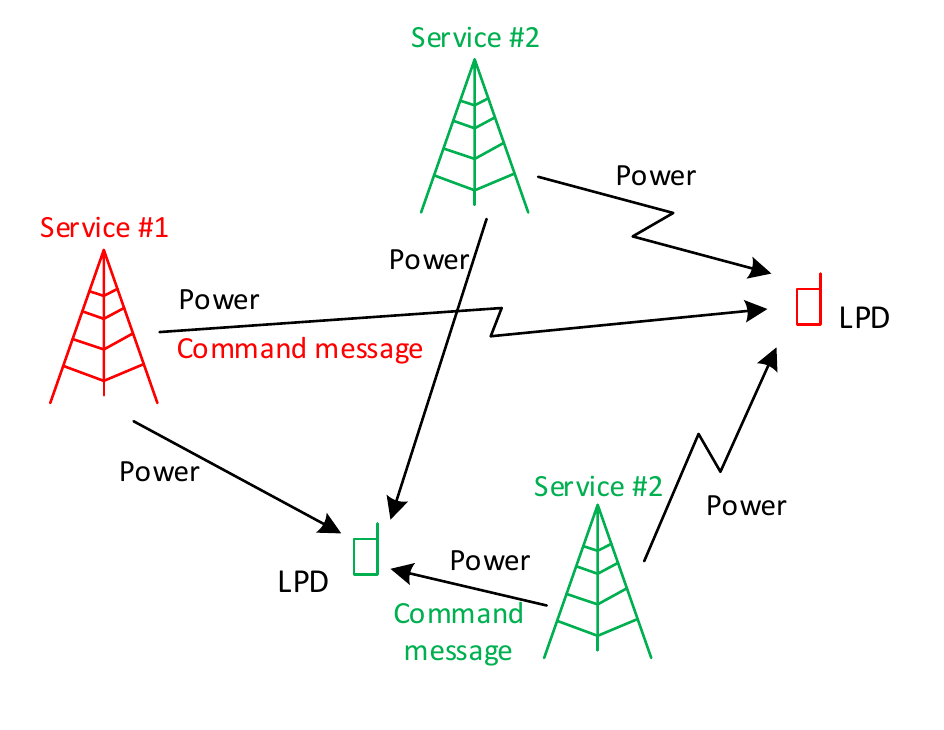}
	\caption{SWIPT-MC system model}
	\label{system} 
\end{figure}

\begin{figure}[!t]
	\centering
	\includegraphics[width=3.5in]{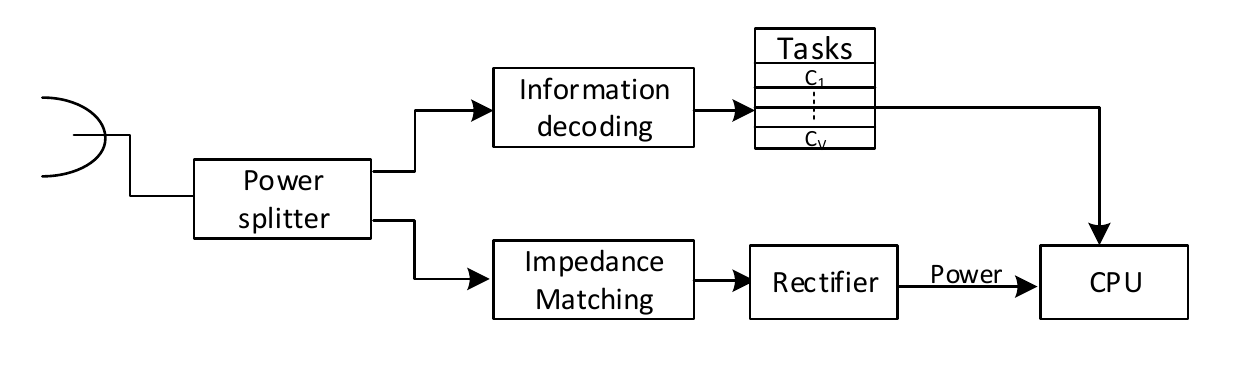}
	\caption{Main components of the LPD}
	\label{LPD} 
\end{figure}

Each PH is equipped with $n_t$ transmit antennas, while the LPDs have $n_r$ receive antennas. Moreover, maximum ratio transmission (MRT) is utilised at the transmitter side while maximum ratio combining (MRC) is implemented at the receiver side. The LPDs embed a SWIPT receiver with power splitting (PS) architecture, \cite{zhang2013}, to decode information and harvest energy simultaneously. As shown in Fig. \ref{LPD}, the received signal is split into two streams of different power levels for decoding and harvesting using a PS ratio $0\le\rho \le 1$. Moreover, it is assumed that the harvested power is a linear function of the average received power. It is worth mentioning that actual RFEHs are nonlinear devices, so that the efficiency of the harvesting process depends on the particular SWIPT waveform adopted and it typically decreases when the average received power decreases. However, for the sake of tractability, we make the hypothesis that the waveform has already been optimised and we consider a constant efficiency $\zeta$ corresponding to the efficiency that the RFEH would have for the minimum harvestable power. In this way, our linear model will provide a lower bound for the harvested power. 

Since an indoor environment is considered, the signal propagation model is composed of distance dependent path-loss, wall blockages and small scale fading. To model the wall blockages, we consider randomly distributed walls following a Manhattan Poisson line process (MPLP) distribution with frequency $\lambda_W$ for both the $x$- and $y$- dimensions. The joint effect of path-loss and wall blockages is given by \cite{akin2018}
\begin{equation}
	l_{W}(r)=
	\begin{cases}
	{\frac{\kappa r^\beta}{K^W}} \quad \quad  \textrm{if} \quad \quad r\ge\kappa^{-1/\beta} \\
	{1} \quad \quad \quad \quad \quad \quad    \textrm{otherwise} ,
	\end{cases}
	\end{equation}
where $r$ is the distance between the PH and the LPD, $\beta$ is the path-loss exponent, $K\in (0,1]$ is the penetration loss and W is a random variable indicating the number of walls between a generic PH and the LDP. $\kappa= \left(\frac{4\pi}{v}\right)^2 $ is the path-loss constant, where $v=c_0/f_c$ is the transmission wavelength, $f_c$ and $c_0$ being the carrier frequency in Hz and the speed of light in m/sec, respectively.

\subsection{Computation Model}

The objective of our MC system is to provide a set of services $\mathcal{S}$, whose generic instance $s$ corresponds to a collection of computation tasks $\mathcal{C}_s$.
To achieve this, the generic PH $p \in \mathcal{P}_s$ transmits a command message with a fixed length of $M$ bits to trigger the execution of a specific task $c\in\mathcal{C}_s$ to process the data gathered locally. Each computation task $c_v, v=1,\cdots,V$ is characterised by $k_v$ logical operations per bit. The primary engine for local computation at the LPD is the CPU, where the energy consumption per logical operation depends on the CPU clock speed. Therefore, under the assumption of low CPU voltage, the energy consumed by the CPU to execute the task $c_v$ can be modelled as \cite{burd1996processor}
\begin{equation}
\label{Eq:EC}
E_C=\sum_{i=1}^{k_vN}{\xi f_i^2},
\end{equation}
in which $N$ is the number of bits encoding the contextual information, $\xi$ is the effective capacitance coefficient and $f_i$ denotes the CPU frequency (i.e. the clock speed) for each CPU cycle $i\in \{1,...,k_vN\}$. 

\section{Rate-Energy Trade-off Analysis}
\label{SG_analysis}

The performance of the SWIPT-MC network will first be illustrated through the achievable trade-offs between the harvested power and the information rate. In this section, we provide an SG framework to investigate the rate-energy trade-off for a generic LPD. 

To achieve this, we consider a typical LPD, indexed by $l=0$ and denoted with LPD$_0$, that is located at the centre of a circular area with radius $R_D$. Typically, the PH experiencing the minimum signal attenuation is assumed to be the serving PH for the LPD$_0$. However, assuming that the LPD$_0$ is only associated with a subset of services $\mathcal{S}_{LPD} \subseteq \mathcal{S}$, it may not be able to answer to the PH's request. In such a case, the PH will initiate a communication with another LPD, thus creating interference to the LPD$_0$. It follows that the serving PH shall belong to the subset of PHs that were allocated to a service $s \in \mathcal{S}_{LPD}$. Therefore, we introduce the concept of hit probability, $q_{hit}=\sum_{s \in \mathcal{S}_{LPD}} q_s$, defined as the probability that a generic PH is allocated to one of the services implemented by the LPD$_0$.
The original PPP, $\Psi$, is then partitioned into two homogeneous PPPs:  $\Psi_{q_{hit}}$ with density $q_{hit}\lambda_{PH}$ and ${\Psi}_{\bar{q}_{hit}}$ with density $\bar{q}_{hit}\lambda_{PH}$, where $\bar{q}_{hit}=1-q_{hit}$. 

In order to analyse the attenuation due to the blockage objects, we follow the procedure originally proposed in \cite{akin2018}, in which homogeneous PPPs are decomposed into the sum of inhomogeneous PPPs, each of which being associated with the probability of experiencing $W$ blockage objects. Hence, we have 
\begin{equation}
\Psi_{q_{hit}}=\sum_{W=0}^{W_{max}}{\Psi_{W,{q}_{hit}}},
\end{equation} 
and 
\begin{equation}
{\Psi}_{\bar{q}_{hit}}=\sum_{W=0}^{W_{max}}{\Psi_{W,\bar{q}_{hit}}},
\end{equation} 
where $W_{max}$ is the number of walls beyond which the attenuation due to blockage makes
	negligible the signal contribution to the rate-energy trade-off. Since the walls are spatially distributed according to an MPLP, the density of the generic PPP $\Psi_{W,q}$ is given by  
\begin{equation}
\lambda_{W,q}\left(r,\theta\right)= q \lambda_{PH}P_{W} \left(r,\theta\right),
\end{equation}
where
\begin{equation}
\begin{aligned}
\label{P_{W}}
P_{W} \left(r,\theta\right)=  \frac{\left(\lambda_{w}r|\cos(\theta)|+\lambda_{w}r|\sin(\theta)|\right)^W }{W!} \exp\left\{-\left(\lambda_{w}r|\cos(\theta)|+\lambda_{w}r|\sin(\theta)|\right)\right\}, \nonumber
\end{aligned}
\end{equation} 
is the probability that a PH located at the point defined by the polar coordinates $(r,\theta)$ experiences the blockage effect of $W$ obstacles. 

Given the partition of the original PPP into elementary elements, the average propagation loss experienced by the serving PH can be expressed as
\begin{equation}
L^{(0)}= \min_{W}\{\min_{n \in \Psi_{W,q_{hit}}}\{l_W(r^{(n)})\}\},
\end{equation}
where $r^{(n)}$ denotes the distance from a generic PH to the LPD$_0$.
It is worth noting that since the choice of the serving PH is associated with the geometry of both the PHs’ spatial distribution and the placement of the walls, the preliminary step enabling the analysis of the performance for the system depicted in Fig. 1 is the characterisation of the stochastic properties of the minimum path-loss, $L^{(0)}$, and the multi-user interference, $\mathcal{I}_{MU}$. According to \cite[Lemma 1]{lu2015stochastic}, the cumulative distribution function (CDF) of $L^{(0)}$ is given by
\begin{equation} \label{eq:FL0}
F_{L^{(0)}}(\alpha)=\Pr\{L^{(0)} \le \alpha\}=1-\prod_{W=0}^{W_{max}} \exp\left\{-\Lambda_{W,q_{hit}}([0,\alpha))\right\},
\end{equation}
where $\Lambda_{W,q_{hit}}([0,\alpha))$ is the intensity of the process $L_{W,q_{hit}}=\left\{l_{W}(r^{(n)}), n \in \Psi_{W,q_{hit}} \right\}$.
	\vspace{0.3cm}
	By noting
	\begin{equation}
	\begin{aligned}
	\label{LambdaW}
	&\Lambda_{W,q}([0,\alpha))=\Pr\left\{\frac{\kappa {(r^{(n)})}^\beta}{K^W} \in[0,\alpha),n \in \Psi_{W,q} \right\}
	\\
	=&\lambda_{PH}\int_{0}^{2\pi}\int_{0}^{\infty}\mathcal{H}\left(\alpha-\frac{\kappa r^\beta}{K^W}\right) P_{W} \left(r,\theta\right)rdrd\theta,
	\end{aligned}
	\end{equation}
 closed-form expression of $\Lambda_{W,q_{hit}}([0,\alpha))$, reported here in \eqref{Lambda_kernel}, can be easily obtained from equation \eqref{LambdaW} by substituting the density of the PPP $\lambda_{PH}$ with $q_{hit} \lambda_{PH}$ \cite[Appendix A]{akin2018}.

\begin{equation}
\begin{aligned} \label{Lambda_kernel}
\Lambda_{W,q}([0,\alpha)) = 
\begin{cases}
&\frac{4\;q\lambda_{PH}}{W!}\sum\limits_{i=0}^{\infty}\frac{(-1)^{i}}{(i)!(i+W+2)}\lambda_{w}^{i+W}\left[\frac{2^{\frac{i+W}{2}} \sqrt{\pi}\,\Gamma(\frac{i+W+1}{2})} {\Gamma(\frac{i+W+2}{2})}-\frac{\sqrt{2}\:_2 F_1\left(\frac{1}{2},\frac{i+W+1}{2},\frac{i+W+3}{2},\frac{1}{2}\right)}{i+W+1}\right]\\
&\times \left(\frac{\alpha K^W}{\kappa} \right)^{\frac{i+W+2}{\beta}} \quad \textrm{if} \; \alpha < \frac{R_D^\beta \kappa}{K^W}\\
&\frac{4\;q\lambda_{PH}}{W!}\sum\limits_{i=0}^{\infty}\frac{(-1)^{i}}{(i)!(i+W+2)}\lambda_{w}^{i+W}\left[\frac{2^{\frac{i+W}{2}} \sqrt{\pi}\,\Gamma(\frac{i+W+1}{2})} {\Gamma(\frac{i+W+2}{2})}-\frac{\sqrt{2}\:_2 F_1\left(\frac{1}{2},\frac{i+W+1}{2},\frac{i+W+3}{2},\frac{1}{2}\right)}{i+W+1}\right]\\
&\times R_D^{i+W+2} \quad \quad \quad \quad \textrm{if} \; \alpha \ge \frac{R_D^\beta \kappa}{K^W}
\end{cases}
\end{aligned}
\end{equation}
where $R_D$ is the maximum distance at which a PH's contribution to the global interference has a non-negligible impact on the achievable rate-energy trade-offs.

After having characterised the control link, we study the other main source of randomness impacting the LPD$_0$ performance, i.e. the multi-user interference, $\mathcal{I}_{MU}$. This can be expressed as 
\begin{equation}
\label{eq:I_MU}
\begin{aligned}
\mathcal{I}_{MU}=&\sum_{W=0}^{W_{max}}\left(\sum_{n\in\Psi^{(\,\backslash 0)}_{W,q_{hit}}}\frac{h^{(n)}}{l_{W}(r^{(n)})}+\sum_{n\in\Psi_{W,\bar{q}_{hit}}}\frac{h^{(n)}}{l_{W}(r^{(n)})}\right),
\end{aligned}
\end{equation}
where $h^{(n)}$ represents the gain of the $n$th interfering link that can be modelled as an exponentially distributed random variable with unit variance \cite[Eq.(38)]{kang2003comparative}. We can notice that $\mathcal{I}_{MU}$ is composed of two parts. The first one represents the interference produced by all the PHs belonging to $\Psi^{(\,\backslash 0)}_{W,q_{hit}}$, while the second one refers to the interference generated by the PH belonging to $\Psi_{W,\bar{q}_{hit}}$.  

Taking advantage of the Probability Generating Functional (PGFL) theorem \cite[Proposition 1.2.2]{baccelli2009}, we find that the processes 
\begin{equation}
L^{(\backslash 0)}_{W,q_{hit}}=\left\{l_{W}(r^{(n)}), n \in \Psi^{(\backslash 0)}_{W,q_{hit}} \right\},
\end{equation} 
and 
\begin{equation}
L_{W,\bar{q}_{hit}}=\left\{l_{W}(r^{(n)}), n \in \Psi_{W,\bar{q}_{hit}} \right\},
\end{equation}  
collecting all the attenuations associated with $\Psi^{(\backslash 0)}_{W,{q}_{hit}}$ and $\Psi_{W,\bar{q}_{hit}}$, are still PPPs with intensities $\Lambda_{W, q_{hit}}([L^{(0)},\alpha))$ and $\Lambda_{W,\bar{q}_{hit}}([0,\alpha))$, respectively. Hence, since $L^{(\backslash 0)}_{W,{q}_{hit}}$ and $L_{W,\bar{q}_{hit}}$ are independent PPPs, from \cite[Proposition 2]{akin2018}, we obtain the CDF of the multi-user interference as
\begin{equation} \label{eq:FIMU}
F_{\mathcal{I}_{MU}}(z;L^{(0)})=\Pr\{\mathcal{I}_{MU} \le z \arrowvert L^{(0)}\}=1/2-\int_0^\infty\frac{1}{\pi \omega}\operatorname{Im}\left\{\textrm{e}^{-j\omega z}\prod_{W=0}^{W_{max}} \Phi_W\left(\omega; L^{(0)}\right)\right\} d\omega,
\end{equation}
where $\Phi_W\left(\omega; L^{(0)}\right)$ is the characteristic function of the interference generated from the PHs associated with $\Psi_{W}=\Psi_{W,q_{hit}}+\Psi_{W,\bar{q}_{hit}}$ and can be expressed as
		\begin{equation}
	\Phi_W\left(\omega; L^{(0)}\right)=\Phi_{W,q_{hit}}\left(\omega; L^{(0)}\right)\Phi_{W,\bar{q}_{hit}}\left(\omega;1\right).
	\end{equation}
		Here $\Phi_{W,q_{hit}}\left(\omega; L^{(0)}\right)$ and $\Phi_{W,\bar{q}_{hit}}\left(\omega;1\right)$, whose expressions are provided in \eqref{Phi_kernel_hit}-\eqref{eq:DEnohit}, are the characteristic functions of the interference generated from the PHs associated with $\Psi^{(\backslash 0)}_{W,q_{hit}}$ and $\Psi_{W,\bar{q}_{hit}}$, respectively and can be calculated as
	\setcounter{equation}{18}
	\begin{equation} \label{eq:Phi}
\Phi_{W,q}\left(\omega; L^{(0)}\right) =\exp\Bigg(\mathbb{E}_{h ^{(n)}}\Bigg\{\int_{L^{(0)}}^\infty \bigg(\exp\bigg(j\omega h^{(n)}/\alpha\bigg)-1\bigg)\times\widehat{\Lambda}_{W,q}([0,\alpha))d\alpha \Bigg\}\Bigg),
\end{equation}
in which $\widehat{\Lambda}_{W,q}([0,\alpha))$ is the first derivative of $\Lambda_{W,q}([0,\alpha))$ with respect to $\alpha$.

\begin{figure*}[!t]
	\setcounter{equation}{14}
\begin{equation}
\begin{aligned}
\label{Phi_kernel_hit}
\Phi_{W,{q}_{hit}}\left(\omega; L^{(0)}\right) = 
\begin{cases}
&\exp\Bigg\{\frac{4\;q_{hit}\lambda_{PH}}{W!}\sum\limits_{i=0}^{\infty}\frac{(-1)^{i}}{(i)!(i+W+2)}\lambda_{w}^{i+W}\left[\frac{2^{\frac{i+W}{2}} \sqrt{\pi}\,\Gamma(\frac{i+W+1}{2})} {\Gamma(\frac{i+W+2}{2})}-\frac{\sqrt{2}\:_2 F_1\left(\frac{1}{2},\frac{i+W+1}{2},\frac{i+W+3}{2},\frac{1}{2}\right)}{i+W+1}\right]
\\
&\times \Delta_{W,{q}_{hit}}\left(\omega;L^{(0)}\right)\Bigg\} \quad \textrm{if} \; L^{(0)} < \frac{R_D^\beta \kappa}{K^W} \\
&1 \quad \quad \quad \quad \quad \quad \quad \quad \quad \quad \textrm{if} \; L^{(0)} \ge \frac{R_D^\beta \kappa}{K^W}
\end{cases}
\end{aligned}
\end{equation}
\hrulefill
\begin{equation}
\begin{aligned}\label{Phi_kernel_nohit}
\Phi_{W,\bar{q}_{hit}}\left(\omega;1\right) = 
&\exp\Bigg\{\frac{4\;\bar{q}_{hit}\lambda_{PH}}{W!\beta}\sum\limits_{i=0}^{\infty}\frac{(-1)^{i}}{(i)!}\lambda_{w}^{i+W}\left[\frac{2^{\frac{i+W}{2}} \sqrt{\pi}\,\Gamma(\frac{i+W+1}{2})} {\Gamma(\frac{i+W+2}{2})}-\frac{\sqrt{2}\:_2 F_1\left(\frac{1}{2},\frac{i+W+1}{2},\frac{i+W+3}{2},\frac{1}{2}\right)}{i+W+1}\right]
\\
&\times \Delta_{W,\bar{q}_{hit}}\left(\omega;1\right) \Bigg\}
\end{aligned}
\end{equation}
\hrulefill

	\begin{equation}
\begin{aligned}\label{eq:DE}
\Delta_{W,{q}_{hit}}\left(\omega;L^{(0)}\right) =&\left(\frac{L^{(0)}K^W}{\kappa}\right)^{\frac{(i+W+2)}{\beta}}\left(1-_2F_1\left(1,-\frac{(i+W+2)}{\beta},1-\frac{(i+W+2)}{\beta},\frac{j\omega}{L^{(0)}}\right)\right) \\
&-R_{D}^{(i+W+2)}\left(1-_2F_1\left(1,-\frac{(i+W+2)}{\beta},1-\frac{(i+W+2)}{\beta},\frac{j\omega K^W}{R_{D}^\beta\kappa}\right)\right)
\end{aligned}
\end{equation}
\hrulefill
	\begin{equation}
	\label{eq:DEnohit}
	\Delta_{W,\bar{q}_{hit}}\left(\omega;1\right) = R_{D}^{(i+W+2)}\left[\frac{\frac{R_{D}^{\beta}\kappa}{K^W}\left(\frac{j}{\omega}\right)\; _2F_1\left(1,1+\frac{(i+W+2)}{\beta},2+\frac{(i+W+2)}{\beta},-\frac{jR_{D}^{\beta}\kappa}{K^W\omega}\right)}{1+\frac{(i+W+2)}{\beta}}-\frac{\beta}{(i+W+2)}\right]
	\end{equation}
	\hrulefill

\vspace*{4pt}
\end{figure*}

The rate-energy trade-off will be analysed through the joint complementary cumulative distribution function (J-CCDF) of the achievable information rate, $R$, and the average harvested power, $Q$: 

\setcounter{equation}{19}
\begin{equation}
F_{c}(R^*,Q^*)=\Pr\{R\geq R^*,Q\geq Q^* \},
\end{equation}
where $R^*\geq 0$ and $Q^*\geq 0$ represent the minimum required information rate and harvested power, respectively.
Assuming that the transmission bandwidth is equal to $B$ and that each PH transmits with average power $P$, we have
\begin{equation}
\begin{aligned}
\label{RQ}
&R = B \log_2{\left(1+\frac{P\,g^{(0)}/L^{(0)}}{P\,\mathcal{I}_{MU} + \sigma^2_n + \sigma^2_c\,/(1-\rho)}\right)},
\\
&Q = \rho \, \zeta \, P \left( \frac{g^{(0)}}{L^{(0)}}+\mathcal{I}_{MU} \right),
\end{aligned}
\end{equation}
where $\rho$ is the power splitting ratio, $\zeta$ is the efficiency of the RFEH, $\sigma^2_n$ denotes the variance of the thermal noise and $\sigma^2_c$ indicates the variance of the noise due to the RF-to-DC conversion of the received signal. The random variable $g^{(0)}$ characterises the power gain of the link between the serving PH and the LPD$_0$, including MRT and MRC, and its probability density function (PDF), which is the largest eigenvalue of the channel matrix, is  given as \cite{maaref2005closed}
\begin{equation}
\label{zeta}
f_{g^{(0)}}(\zeta)=\mathcal{K}_{m,n}\sum_{s=1}^{m}\sum_{t=n-m}^{(n+m-2s)s}a_{s,t}\zeta^t\exp(-s\zeta),
\end{equation}
where $\mathcal{K}_{m,n}=\prod_{i=1}^{m}((m-i)!(n-i)!)^{-1}$ is a normalizing factor with $m=\min(n_{t},n_{r})$ and $n=\max(n_{t},n_{r})$, and the coefficients $a_{s,t}$ can be computed using \cite[Algorithm 1]{maaref2005closed}.

Once we have all the statistical characterisations of $L^{(0)}$, $\mathcal{I}_{MU}$, and $g^{(0)}$ expressed in equations \eqref{eq:FL0}, \eqref{eq:FIMU}, and \eqref{zeta}, respectively, we can take advantage of the convenient reformulation of J-CCDF which has been proposed in \cite{thanh2016mimo}. It amounts to computing the probability that the multi-user interference belongs to an interval for which a minimum SINR $\gamma$ can be achieved conditioned to a minimum amount of received power $q_*$. Hence, after some manipulations, using \eqref{RQ} we get
\begin{equation}
\begin{aligned}
\label{F_C}
&F_{c}(R^*,Q^*)=\mathbb{E}_{L^{(0)}}\left\{\int_{(\mathcal{T}_*/P)L^{(0)}}^{+\infty}F_{\mathcal{I}_{MU}}\left({\frac{x\gamma}{L^{(0)}}}-\frac{\sigma_{*}^2}{P}\middle\arrowvert L^{(0)}\right)f_{g^{(0)}}(x)dx\right\}
\\
&-\mathbb{E}_{L^{(0)}}\left\{\int_{(\mathcal{T}_*/P)L^{(0)}}^{+\infty}F_{\mathcal{I}_{MU}}\left({\frac{-x}{L^{(0)}}}+\frac{q_{*}}{P}\middle\arrowvert L^{(0)}\right)f_{g^{(0)}}(x)dx\right\},
\end{aligned}
\end{equation}
\vspace{0.2cm}
where  $\sigma_{*}^2=\sigma_{n}^2+ \frac{\sigma_{c}^2}{1-\rho}$, $q_{*}=\frac{Q^*}{\rho\zeta }$, $\mathcal{T}_{*}=\frac{q_{*}+\sigma_{*}^2}{\gamma+1}$ and $\gamma=1/\left(2^{R^*/B}-1\right)$.
\vspace{0.1cm}
Eventually, by substituting \eqref{eq:FL0}, \eqref{eq:FIMU}, and \eqref{zeta} into \eqref{F_C}, the J-CCDF $F_c(R^*,Q^*)$ is obtained through the expression \cite{thanh2016mimo}
\begin{equation} \label{JCCDF}
F_{c}(R^*,Q^*)=\mathcal{K}_{m,n}\sum_{s=1}^{m}\sum_{t=n-m}^{(n+m-2s)s}a_{s,t}\left(J_{s,t}^{(1)}-J_{s,t}^{(2)}\right).
\end{equation}
The functions $J_{s,t}^{(1)}$ and $J_{s,t}^{(2)}$ are defined as
\begin{align}
\label{J_{s,t}^{(1)}}
J_{s,t}^{(1)}=\int_{0}^{\infty}\int_{0}^{\infty} &\frac{1}{\pi \omega} \operatorname{Im}\Bigg\{\exp\left(-j\omega\frac{q_{*}}{P}\right)  {\left(s-\frac{j\omega}{y}\right)}^{-(1+t)}
\Gamma\left(1+t,\frac{\mathcal{T}_{*}}{P}(sy-j\omega)\right) \nonumber \\
&\times\prod_{W=0}^{W_{max}} \Phi_W\left(\omega;y\right)\Bigg\}\left(\widehat{\Lambda}_{q_{hit}}([0,\alpha))\exp\left\{-\Lambda_{q_{hit}}([0,\alpha))\right\}\right)d\omega dy,
\end{align}
\begin{align}
\label{J_{s,t}^{(2)}}
J_{s,t}^{(2)}=\int_{0}^{\infty}\int_{0}^{\infty}&\frac{1}{\pi \omega}\operatorname{Im}\Bigg\{\exp\left(j\omega\frac{\sigma_{*}^2}{P}\right)  {\left(s+\frac{j\omega \gamma}{y}\right)}^{-(1+t)}
\Gamma\left(1+t,\frac{\mathcal{T}_{*}}{P}(sy+j\omega \gamma)\right) \nonumber \\
&\times\prod_{W=0}^{W_{max}} \Phi_W\left(\omega; y\right)\Bigg\}\left(\widehat{\Lambda}_{q_{hit}}([0,\alpha))\exp\left\{-\Lambda_{q_{hit}}([0,\alpha))\right\}\right)d\omega dy,
\end{align}
with  $\Lambda_{q_{hit}}([0,\alpha))=\sum_{W=0}^{W_{max}}\Lambda_{W,q_{hit}}([0,\alpha))$, and where $\widehat{\Lambda}_{q_{hit}}([0,\alpha))$ denotes the first derivative of $\Lambda_{q_{hit}}([0,\alpha))$ with respect to $\alpha$.

\section{SWIPT based Real time Computation}
\begin{figure}[!t]
	\centering
	\includegraphics[width=3.7in]{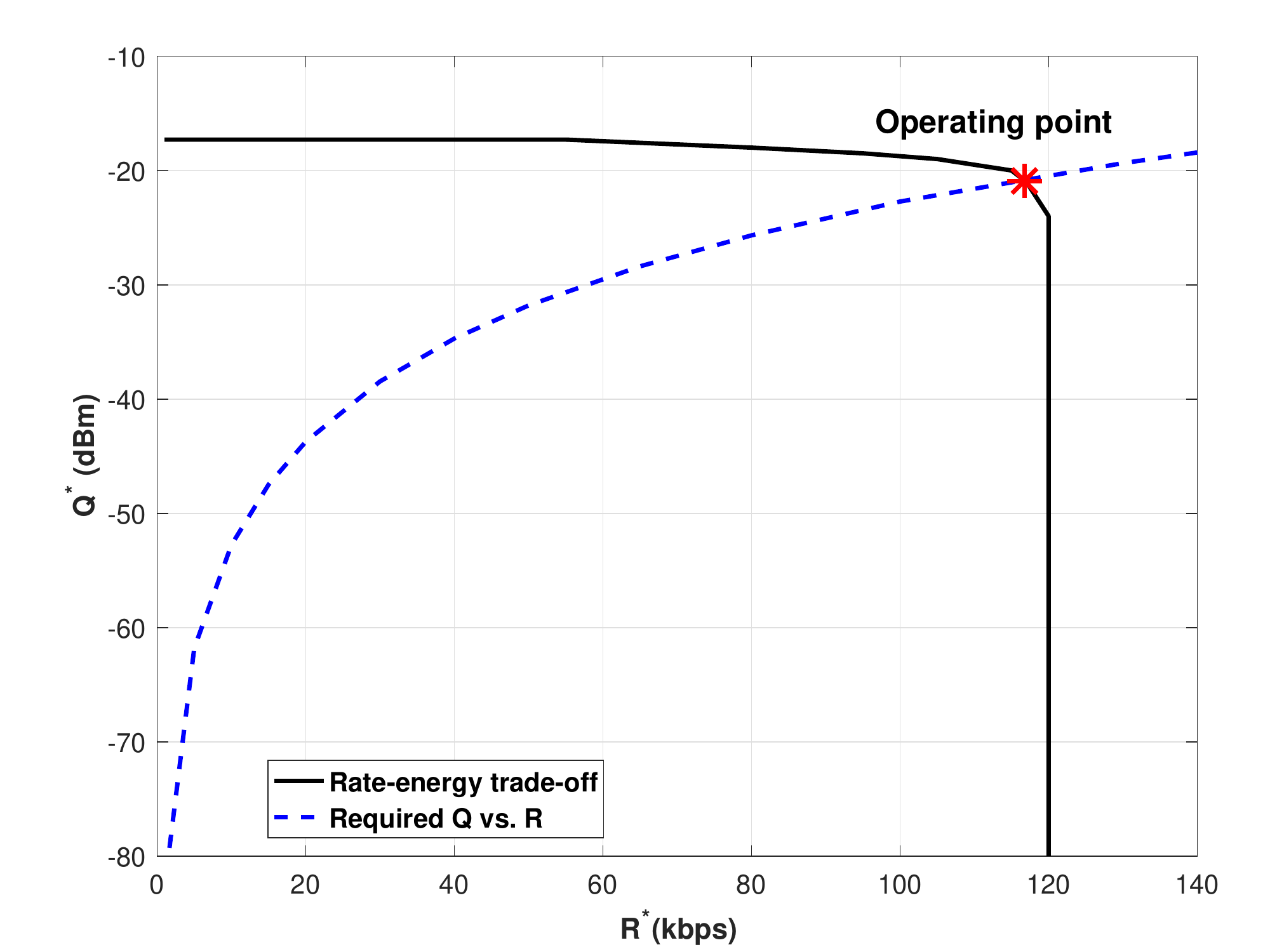}
	\caption{Illustration of the rate-energy trade-off, the required harvesting power versus data rate and 
		their intersection. }
	\label{example}
\end{figure}

The J-CCDF analysis proposed in Section \ref{SG_analysis} provides a comprehensive understanding of the rate-energy trade-offs that are achievable with a given probability. For illustrative purposes, an example of trade-off curve is given in Fig.\ref{example}. It can be observed that each curve contains an entire set of available choices for the network designer, where the selection of a particular operating point depends on the application targeted. 

In this work, we aim at analysing a real-time SWIPT-MC system whose objective is to harvest enough energy to support the CPU chore in processing the $N$ bits of local data at a given rate. As already stated in the system model, the LPDs' activities are remotely controlled by the PHs through control messages which determine the tasks to be executed. Upon decoding the $M$ bits of the command message, the LPD exploits the harvested energy to perform the computation tasks requested and the triggered tasks must be executed prior to the reception of the next message. In other words, the information rate of the wireless link determines the amount of tasks to be executed per unit of time. At the same time, larger values of rate correspond to smaller values of energy harvested. From \eqref{Eq:EC}, this amounts to saying that the CPU must perform less cycles or reduce its clock speed. Clearly, this causes a conflict between communication and computation performance. In the remainder of this section we will describe a procedure to determine the best compromise between communication and computation for a given J-CCDF.

The adopted CPU energy consumption model assumes that the LPD can adjust the CPU frequencies, $f_{i}$, for each CPU cycle $i\in\{1,...,N_{cycle}\}$ by applying dynamic voltage and frequency scaling (DVFS) techniques \cite{mach2017}. Assuming that a service $s$ is associated with $V$ tasks and that each control message triggers $T$ computation tasks, e.g. $c_{s,1}, c_{s,2}, \cdots, c_{s,T}$, with $T \le V$, then the number of CPU cycles will be equal to $N_{cycle}=kN$, with $k=\sum_{v=1}^T{k_v}$ being the total number of logical operations per processed bit. 
Now, the computation capacity can be determined by optimising the clock frequencies to minimise the CPU energy consumption. This problem can be formulated as \cite{wang2017joint}
\begin{equation} \label{P1} 
\begin{aligned}
&\underset{f_i}{\text{Minimize}}
& &  E_C=\sum_{i=1}^{kN}\xi{f_i}^2\\
& \text{subject to} 
& &  \sum_{i=1}^{kN}\frac{1}{f_i}\leq\frac{M}{R}\\
&
&& f_i\in(0,f_{max}]
\end{aligned}
\end{equation}
where the feasibility region is obtained when 
\begin{equation}
R \leq \frac{f_{max}M}{kN},
\end{equation}
with $f_{max}$ representing the maximum CPU clock speed. It can easily be found that the optimal point is achieved when all the frequencies are equal to $f_i=\frac{kNR}{M}, \forall{i}$, which, in turn, gives a minimum energy consumption per received control message
\begin{equation}
\label{QR}
E_c = \xi\frac{R^2(kN)^3}{M^{2}}.
\end{equation}
Hence, the minimum power needed to support real-time operations is given by
\begin{equation}
\label{QR}
Q = \frac{E_c R}{M} = \xi\frac{(kNR)^3}{M^{3}}.
\end{equation}
The above relation will determine the choice of the LPD's operating point. For example, the trade-off curve in Fig. \ref{example} includes all the possible $(R,Q)$ pairs with $F_{c}(R^*,Q^*)=0.75$. The relation between the information rate and the required harvested power for computing $N=0.6$ kbits with $k=20$ operations per bit and $M=32$ bits is also plotted in this figure. As indicated, the intersection of these two curves determines the rate-energy trade-off enabling real-time computation. 

Finally, we can define the outage probability of the real-time SWIPT-MC system as a function of the targeted information rate as
\begin{equation}
\label{outage}
P_{out} = 1-F_{c}\left(R,\xi\frac{(kNR)^3}{M^{3}}\right).
\end{equation}
Interestingly, the information rate can also be interpreted as the rate at which the local data must be updated before initiating a new computation. Hence, \eqref{outage} provides also insights into the maximum task complexity given the refresh rate of the data gathered at the LPD.

\section{Numerical Results}
In this section, we illustrate the joint effect of the network densification and the propagation environment on the performance of the proposed SWIPT-MC system. Moreover, we validate our analytical findings by means of Monte Carlo simulations.

\subsection{Setup}
The parameters are set as follows unless otherwise specified. A circular area with radius of  $R_D=60$m is considered. The PHs are randomly distributed using a PPP  with a density of $\lambda_{PH}=1/(\pi d_{PH}^2)$, where $d_{PH}=3$m is half of the average minimum distance between PHs. The average transmit power is set to $1$W, i.e. $P=30$dBm. The signal bandwidth is assumed to be $B=200$kHz centred around a frequency of $f_c=2.1$GHz. The frequency of the walls is also fixed to $\lambda_w=0.03$. At the LPDs, the variance of the noise due to the RF to DC conversion is set to $\sigma_c^2=-70$dBm and the thermal noise variance is given by $\sigma_n^2=-174+10\log_{10}(B)+\mathcal{F}_{n}$, where $\mathcal{F}_{n}=10$dB is the noise figure. The effective capacitance coefficient is set to $\xi=10^{-28}$, the energy harvesting efficiency is assumed to be $\zeta=0.8$. Here it is worth mentioning that in \cite{akin2018}, the authors have obtained the same value of maximum achievable information rate for all values of power splitting ratio $\rho$ due to the fact that, with the level of densification required to receive a total power belonging to the microwatt region, the system is essentially interference limited and the SINR does not depend on realistic levels of noise. By concluding that the power splitting ratio must be as large as possible and its fine tuning is of limited interest in the design of ultra-dense SWIPT networks, the power splitting factor is set to $\rho=0.99$. The number of antennas at the LPD is $n_r=2$ and the number of transmit antennas is set to $n_t=4$. The length of  the control messages is fixed to $M=32$bits and the bits of local data at the LPD are assumed to be $N=0.6$kb. Moreover, the path-loss exponent is assumed to be $\beta=2.5$ and the penetration loss $K$ is $-10$dB per crossed wall according to \cite[Table 3]{series2012propagation}.

\subsection{Results}
\subsubsection{Validation of the analytical findings}

\begin{figure}[!t]
	\centering
	\includegraphics[width=3.7in]{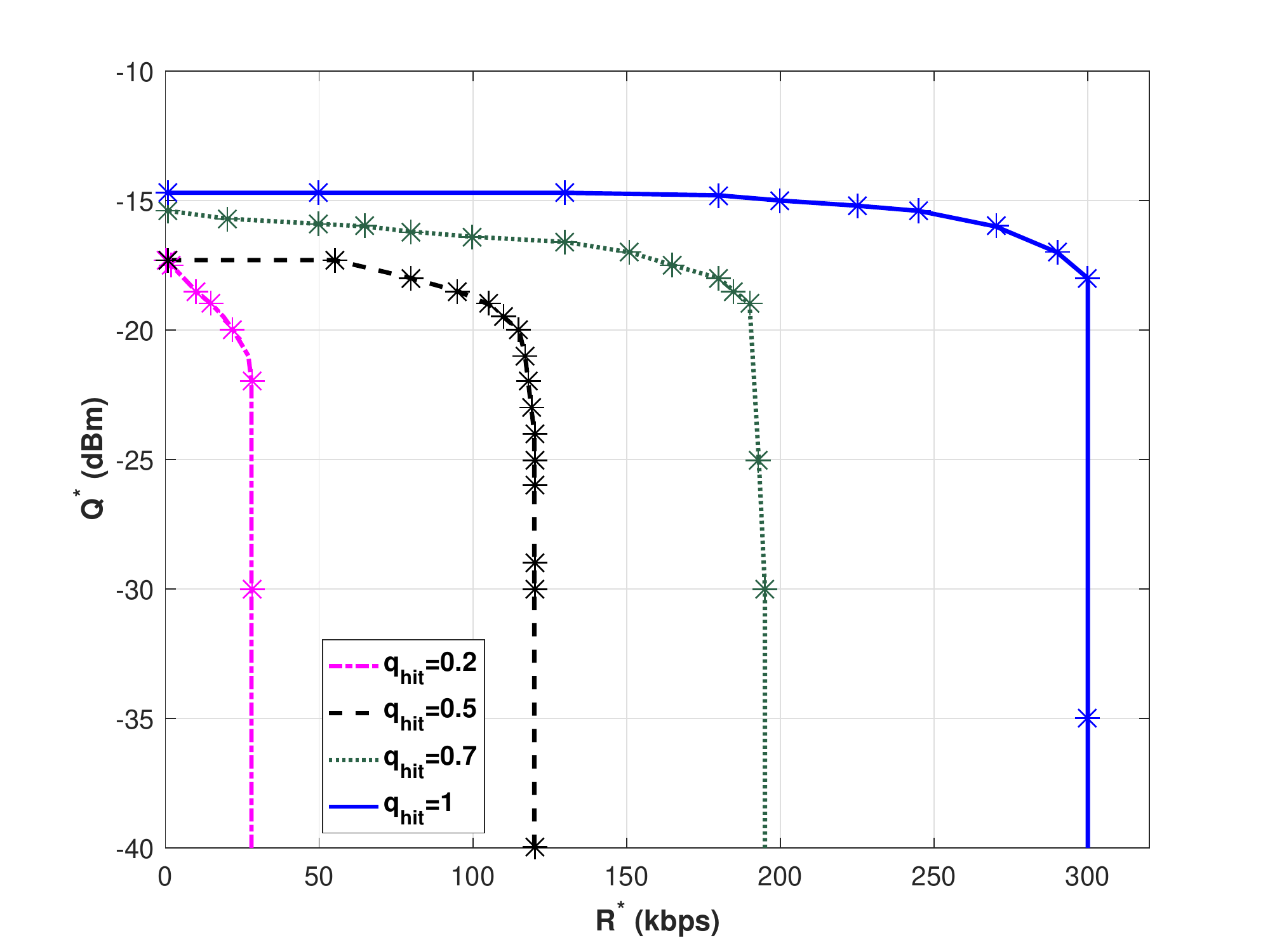}
	\caption{Rate-energy trade-off: theoretical results and Monte Carlo simulations for different values of $q_{hit}$.}
	\label{validate} 
\end{figure}

Fig.\ref{validate} shows the trade-off between the information rate and the harvested power for several values of $q_{hit}$ when $F_c(R^*,Q^*)=0.75$. In order to validate our analysis, both theoretical results (solid lines) and Monte Carlo simulations (markers) are reported. An almost perfect match between the Monte Carlo simulations and the analytical curves is observed. Besides, it can be observed how the hit probability impacts the rate-energy trade-off. In fact, when the hit probability increases, more PHs will be allocated to one of the services implemented by the LPD$_0$. As a result, the probability of experiencing a relatively moderate signal attenuation between the PH and the LPD$_0$ will be larger, which, in turns will give rise to higher values of signal to interference plus noise ratio (SINR) at the SWIPT receiver. For instance, by increasing $q_{hit}$ from $0.2$ to $1$, the information data rate increases from $28$Kbps to $300$Kbps for $-25$dBm of harvested power. From Fig.\ref{validate} we can also observe that the maximum harvested power increases when $q_{hit}$ increases. This is mainly due to the MRT at the serving PH that focuses most of the power at the LPD$_0$ receive antennas. For example at $22$kbps of information rate, with $q_{hit}=1$, the power harvested is $5.3$dB greater than the obtained for $q_{hit}=0.2$. 

\subsubsection{Operating point vs task complexity}

\begin{figure}[!t]
	\centering
	\includegraphics[width=3.7in]{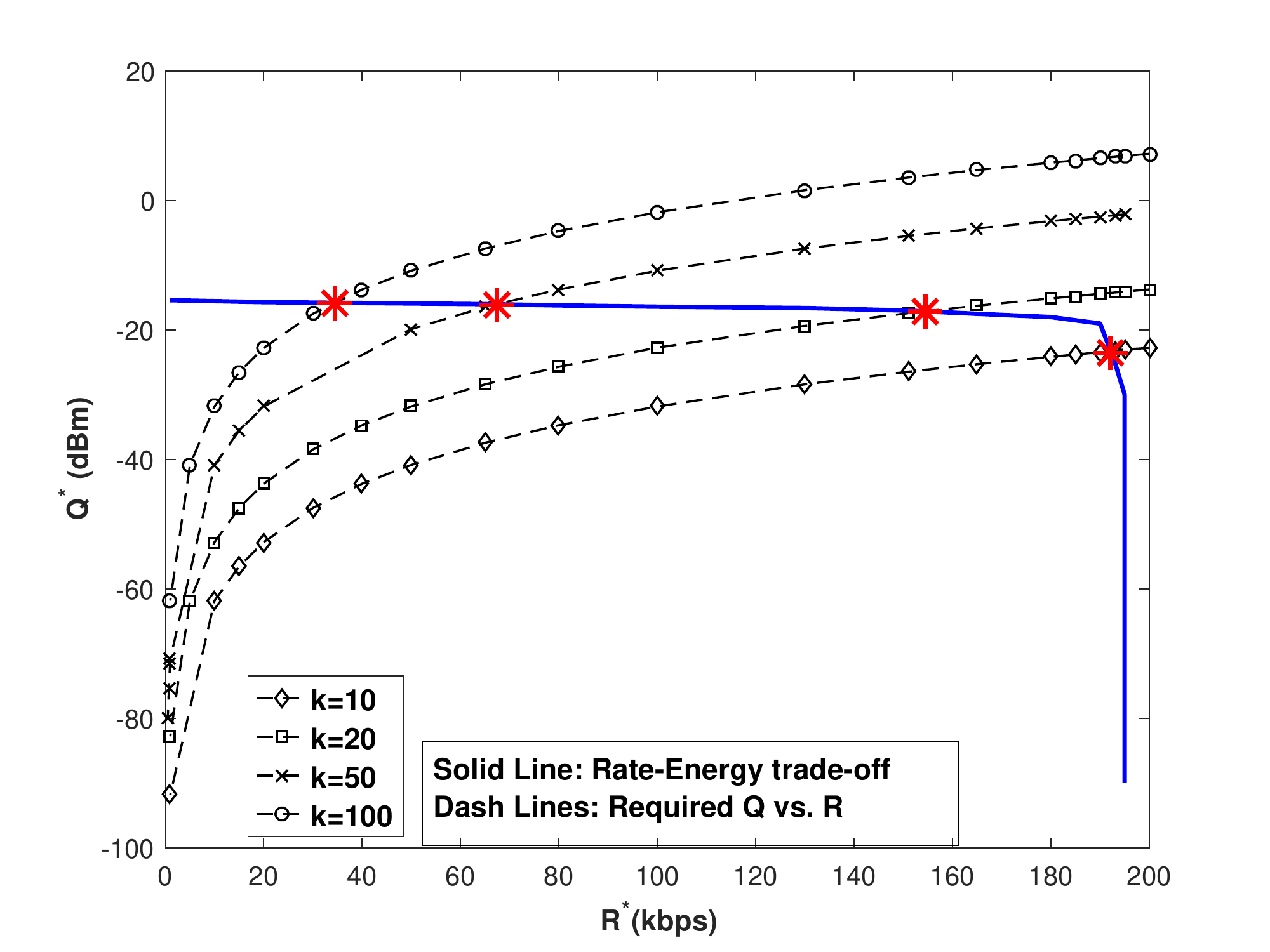}
	\caption{Operating points on the rate-energy trade-off for different values of $k$.}
	\label{operation} 
\end{figure}
Fig.\ref{operation} indicates the operating points associated with different computation loads (expressed in number of logical operations per bit processed). The number of logical operations per bit is set to $k=10,20,50,100$, $F_c(R,Q)=0.75$ and $q_{hit}=0.7$. Not surprisingly, for fixed information rates, more harvested power is required for performing more complex tasks. It can be observed that if $k=10$ operations per bit are required we need $Q^*=-23.5$dBm of power to perform $R^*/M=192/32=6$ kilotasks/second. However, if $k=100$, we can only execute $1$ kilotasks/second by using $Q^*=-15.8$dBm of harvested power.

\subsubsection{Outage probability}
\begin{figure}[!t]
	\centering
	\includegraphics[width=3.7in]{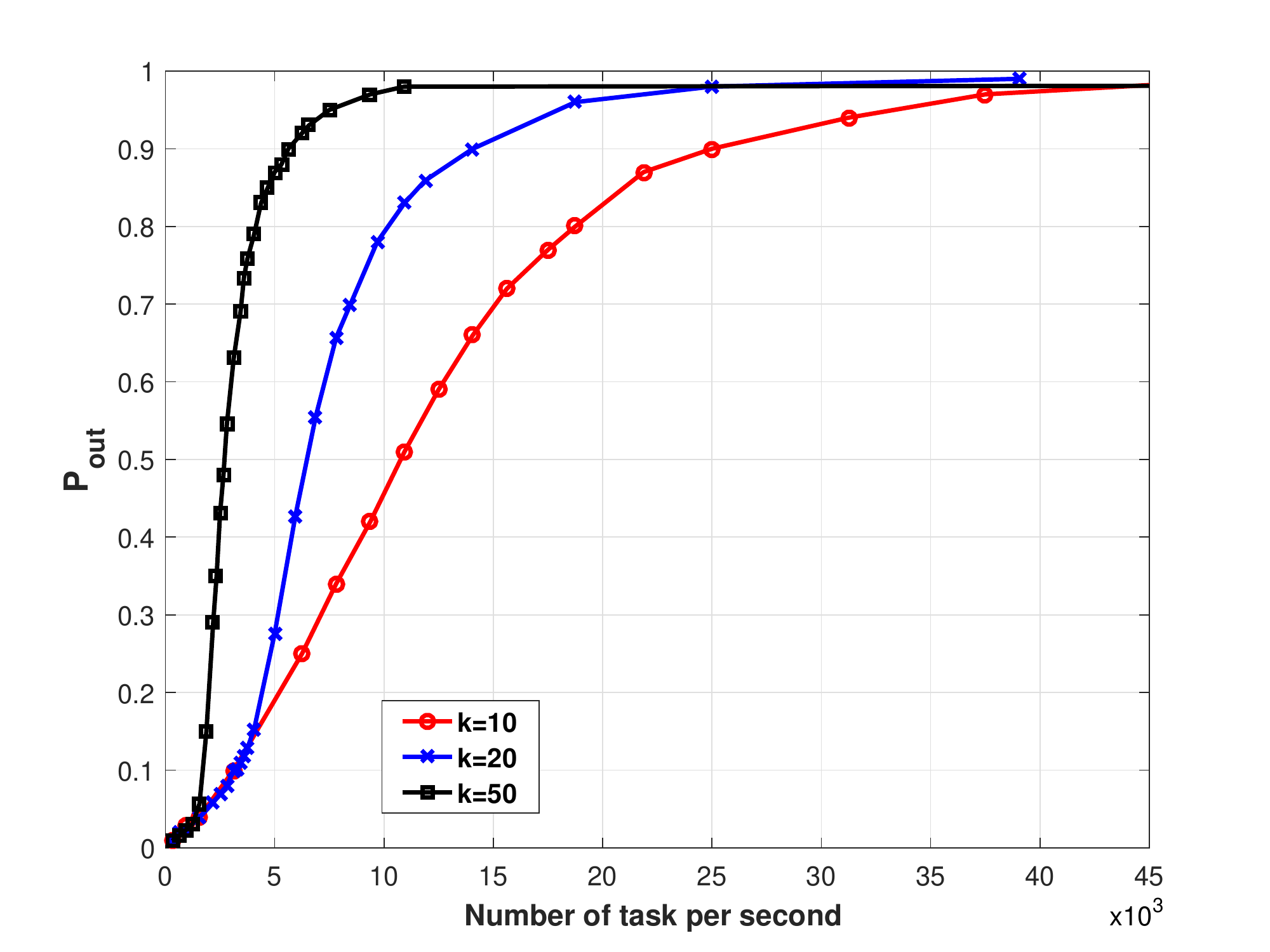}
	\caption{Outage probability versus the number of tasks per second for different values of $k$.}
	\label{P_out_task_sec_k} 
\end{figure}

\begin{figure}[!t]
	\centering
	\includegraphics[width=3.7in]{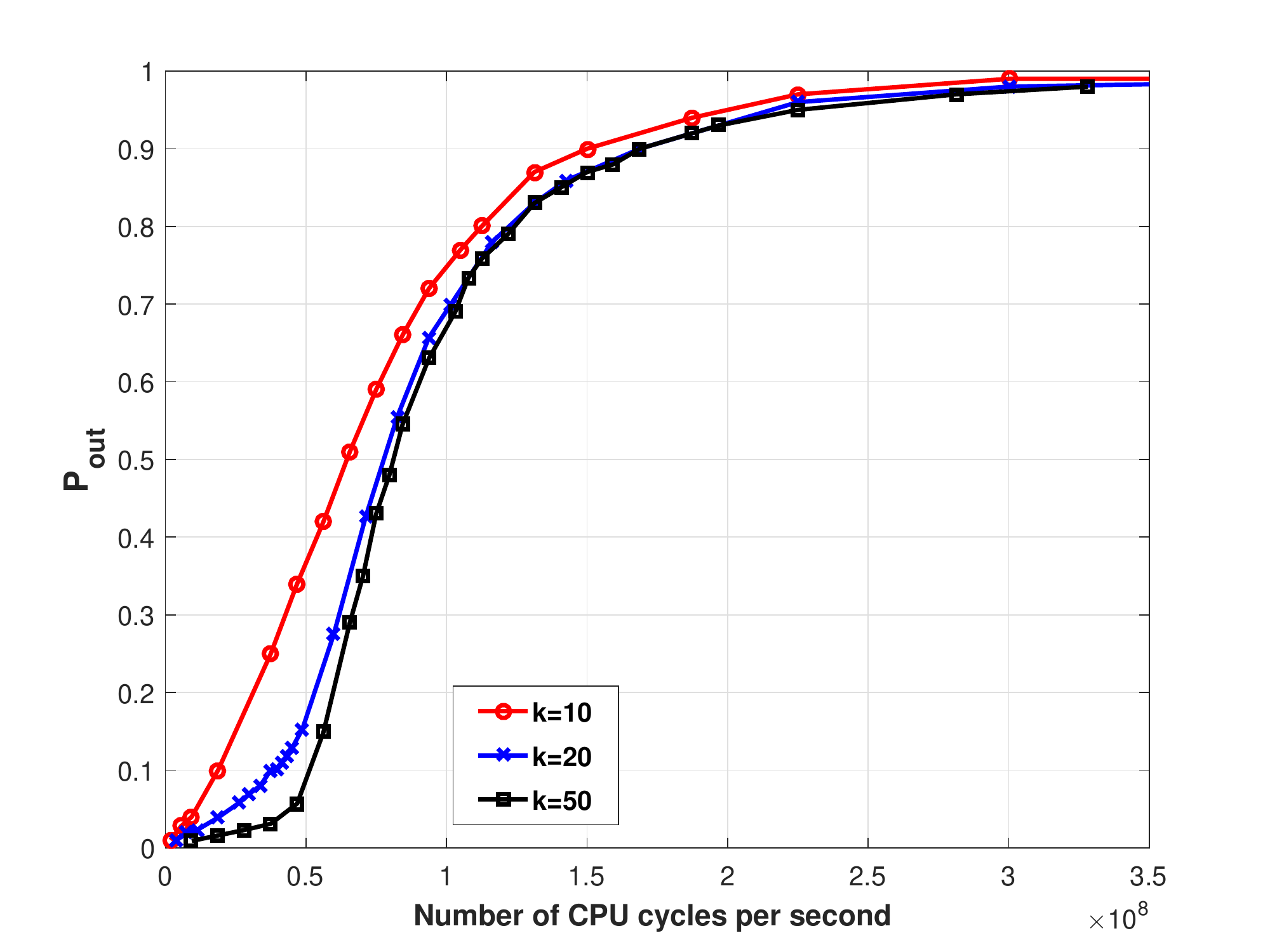}
	\caption{Outage probability versus the number of CPU cycles per second for different values of $k$.}
	\label{P_out_ope_sec_k} 
\end{figure}

In this subsection, we analyse the outage probability of the SWIPT-MC system. 
Fig. \ref{P_out_task_sec_k} illustrates the outage probability against the number of tasks per second to be executed for $k=10,20,50$. It is apparent that, for the same outage probability, the number of tasks per second decreases when the number of operations per bit increases. As an example, with a fixed outage probability of $P_{out}=0.4$, we can perform $9$ kilotasks/second with $k=10$ operations per bit, while for $k=20$ and $k=50$ only $5$ and $2$ kilotasks/second can be executed, respectively. However, as shown in Fig.\ref{P_out_ope_sec_k}, executing more tasks per second does not necessarily mean that we better utilise the CPU.
In fact, from Fig. \ref{P_out_ope_sec_k} we observe that for a fixed number of CPU cycles per second, $\frac{kNR}{M}$, the outage probability decreases when $k$ increases.  This is mainly due to the fact that the data rate is interference limited, so it is not always possible to perform low complexity tasks at a very high rate. Therefore, it can be more efficient to perform more complex tasks at low rate to optimise the CPU usage. Stated otherwise, it would be more efficient to trigger a large number of elementary tasks ($k_v$) with the same command message rather than sending command messages at a very high rate to select less complex tasks.  

\begin{figure}[!t]
	\centering
	\includegraphics[width=3.7in]{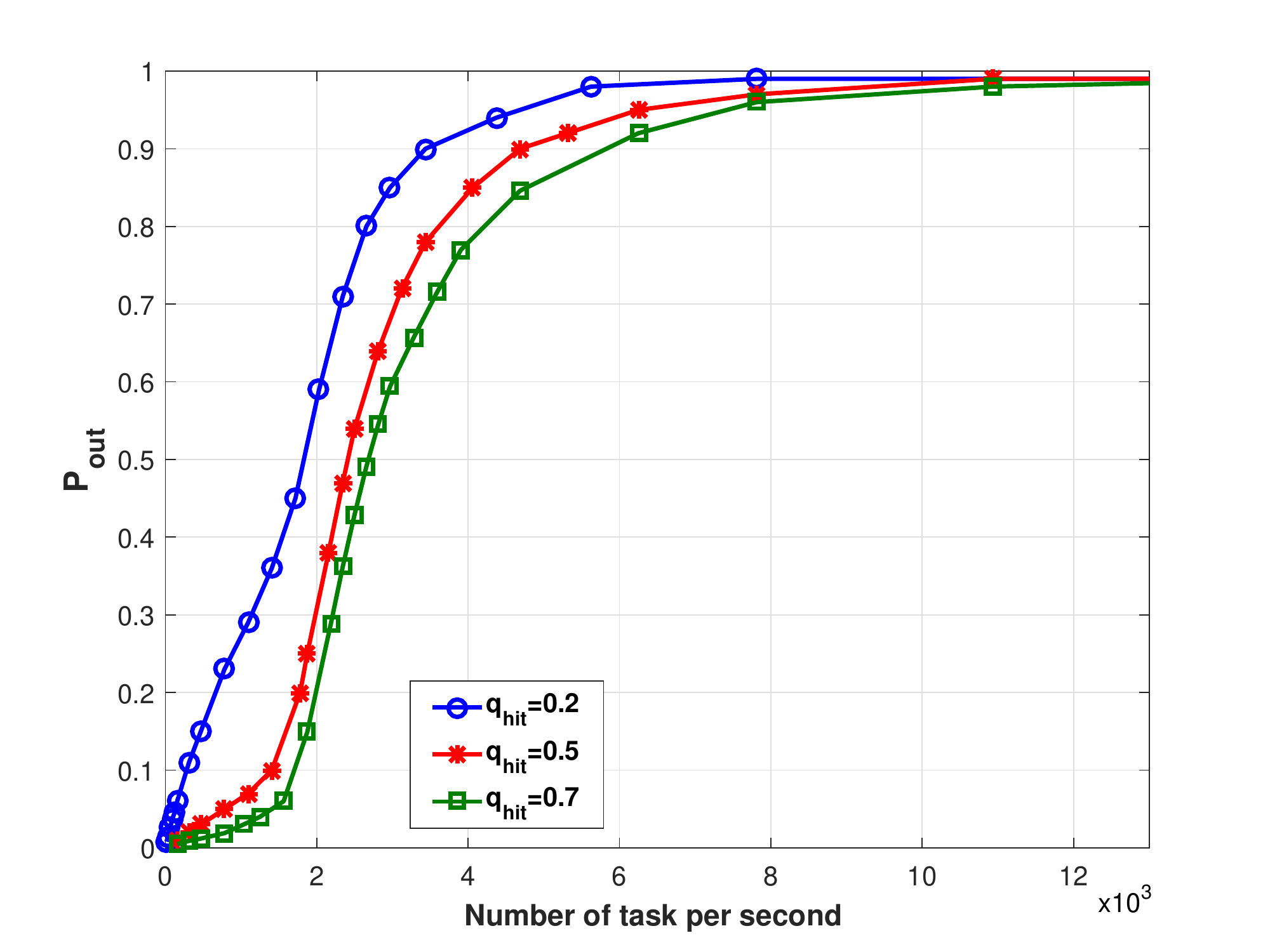}
	\caption{Outage probability versus number of tasks per second for different values of $q_{hit}$.}
	\label{P_out_task_sec_q} 
\end{figure}

\begin{figure}[!t]
	\centering
	\includegraphics[width=3.5in]{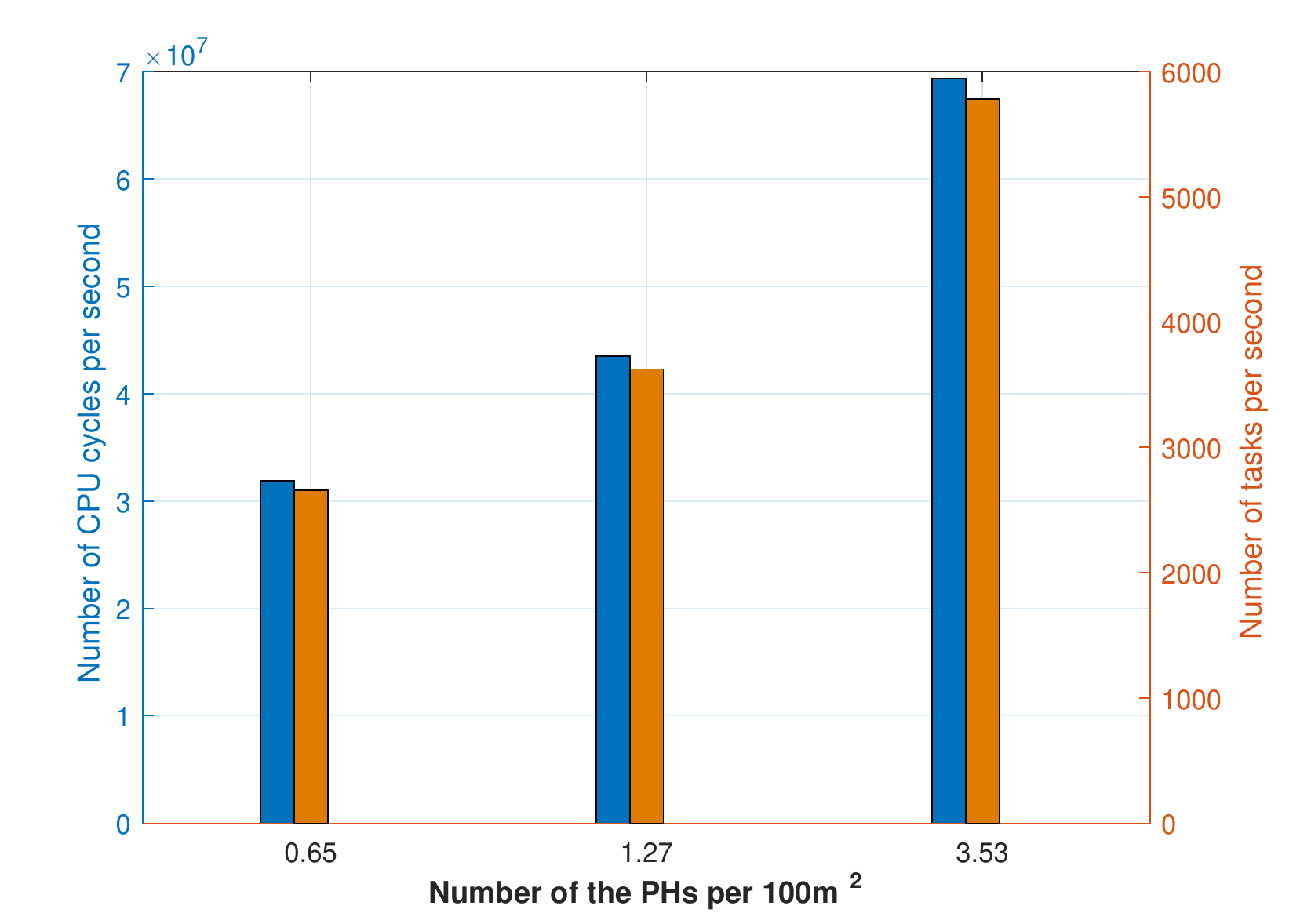}
	\caption{Effect of the PHs' densities for $k=20$ operations per bit.}
	\label{BSeffect_20} 
\end{figure}

\begin{figure}[!t]
	\centering
	\includegraphics[width=3.5in]{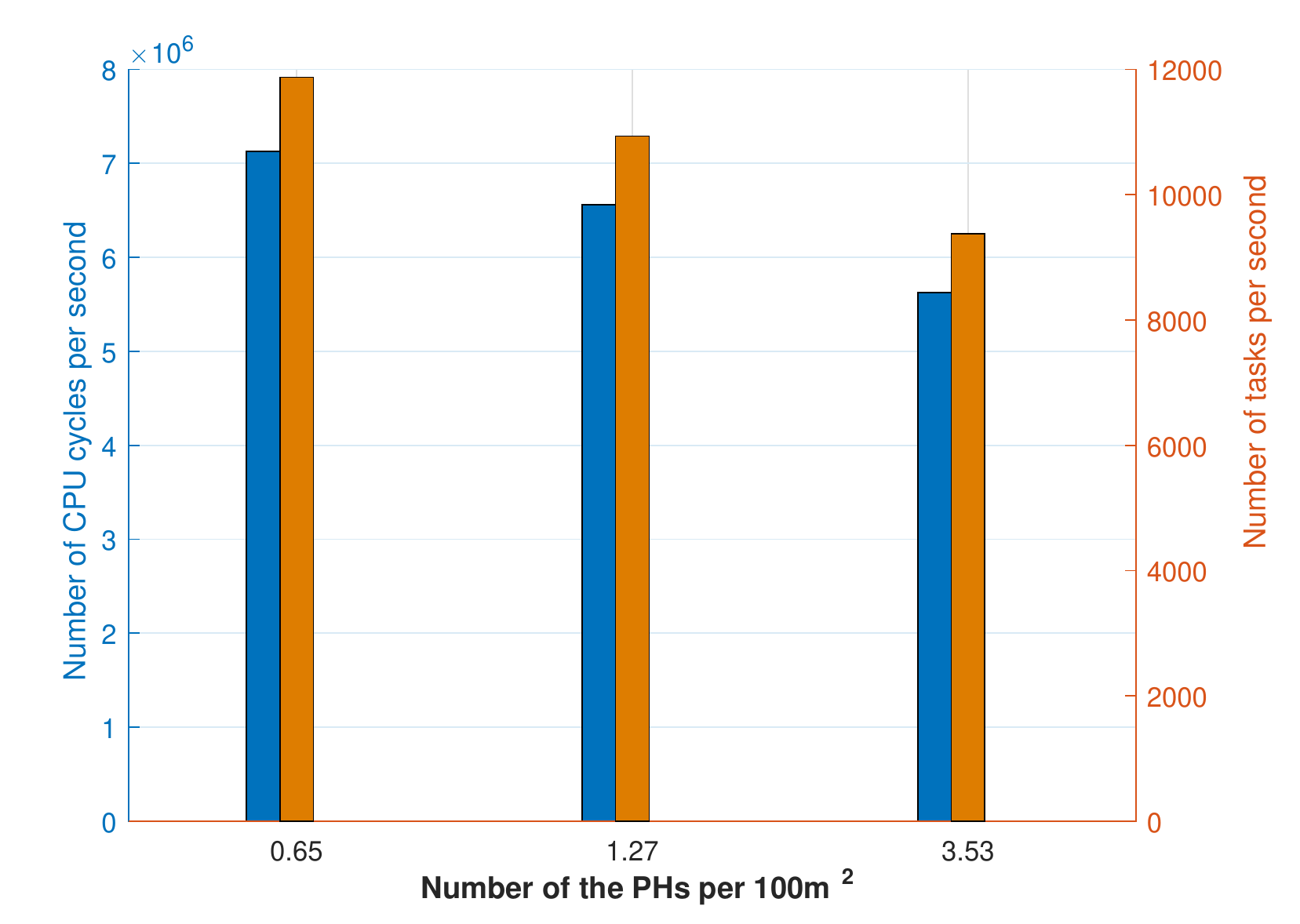}
	\caption{Effect of the PHs' densities for $k=1$ operation per bit.}
	\label{BSeffect_1} 
\end{figure}

In Fig. \ref{P_out_task_sec_q} we investigate the effect of $q_{hit}$ on the outage probability. As illustrated, the number of tasks per second improves when $q_{hit}$ increases. In other words, if the LPDs are more versatile, i.e. they can be associated with more than one service, the SWIPT-MC system will have better performance.
Finally, the effect of network densification is made clear in figures \ref{BSeffect_20} and \ref{BSeffect_1}. Fig. \ref{BSeffect_20} shows the total number of CPU cycles per second and the number of tasks per second for different values of $d_{PH}=3,5,7$m, $q_{hit}=1$ and $k=20$. In this case, network densification is helpful in increasing the SWIPT-MC system performance. However, if only $k=1$ logical operation per bit is considered (e.g. a simple comparison of the local data with a reference value), the CPU is better utilised for a less dense network. This can be explained by the fact that each task consumes very little power and that the limiting factor is not the harvested power but the rate at which the tasks can be performed. It follows that it would be better to have a less dense network to reduce the multi-user interference and increase the information rate at the expense of the harvestable power level.

\section{Conclusion}
In this paper, we proposed a model for real-time MIMO SWIPT-MC systems operating in an indoor environment. We analysed the considered system by making use of an SG framework, providing a comprehensive understanding of the rate-energy trade-off on a service basis. We then investigated the connection between the rate-energy trade-off and the computation capacity of the LPDs. In addition, we provided the optimal operating points guaranteeing real-time processing of the locally generated data for given task complexities. Moreover, the outage probability of the SWIPT-MC system was studied for different network setups. The numerical results show that the best CPU utilisation is obtained when more complex tasks are performed at a lower rate instead of executing less complex tasks at a higher rate. Finally, the effect of network densification on the achievable number of CPU cycles and tasks per second was presented, showing that the optimal level of densification depends on the targeted task complexity. Our theoretical findings were validated through Monte Carlo simulations.

In the setup used in this paper it was assumed that all messages will trigger a deterministic number of logical operations per bit. An improvement on this model would be to have a message dependent stochastic model for the number of CPU cycles. Such a setup will be investigated in a future work. Another potential extension of this work is to consider a SWIPT-based Mobile Edge Computing (MEC) system, in which the computation tasks can be outsourced to a MEC server. 
\section*{Acknowledgement}
This work was supported by F.R.S.-FNRS under the EOS program (EOS project 30452698) and by INNOVIRIS under the COPINE-IOT project.




%
\bibliographystyle{IEEEtran}
\bibliography{Arxiv_Version}
%



%




\end{document}